%% file: main.tex
\documentclass[nonacm,sigconf]{acmart} 
\usepackage[utf8]{inputenc} 

\AtBeginDocument{}
\input{_rights_management}
\input{_aliases}

\begin{document}
\title{\oursystem: Accelerating Human-Centered AI + XR Innovation} 
\subtitle{Preliminary White Paper, September 2025}

\input{_author_list}
\input{0_abstract}
\input{_css}
\input{_keywords}
\input{0_teaser}

\maketitle

\input{1_introduction}
\input{2_related_work}
\input{3_design_space}

\input{5_applications}

\input{7_future_work}
\input{8_conclusion}

\bibliographystyle{ACM-Reference-Format}
\bibliography{99_references}

\end{document}

%% file: _rights_management.tex
\setcopyright{none}
\copyrightyear{2025}
\acmYear{2025}
\acmConference[arXiv '25]{XR Labs Preliminary White Paper}{September 2025}{}
\acmBooktitle{XR Labs Preliminary White Paper}

%% file: _aliases.tex
\usepackage{xspace}
\newcommand{\oursystem}{\textsc{XR Blocks}\xspace}
\newcommand{\systemname}{\textsc{XR Blocks}\xspace}

\usepackage{amsmath}
\usepackage{enumitem}
\usepackage{subfigure}
\usepackage{fontawesome5}

\makeatletter
\newcommand{\github}[1]{%
   \href{#1}{\faGithubSquare}%
}
\makeatother
\usepackage{xcolor}
\usepackage{listings}

\newif \ifdraft \drafttrue   

\definecolor{whatblue}{HTML}{eaf6fc} 
\definecolor{howorange}{HTML}{ffe9d7}
\definecolor{lightgrey}{HTML}{E6E7E9}

%% file: _author_list.tex
\settopmatter{printacmref=false}
\author{David Li$^*$, Nels Numan$^\dagger$, Xun Qian$^\dagger$, Yanhe Chen$^\dagger$, Zhongyi Zhou$^\dagger$,  Evgenii Alekseev, Geonsun Lee,}
 \author{Alex Cooper, Min Xia, Scott Chung, Jeremy Nelson, Xiuxiu Yuan, Jolica Dias, Tim Bettridge,} 
  \author{Benjamin Hersh, Michelle Huynh, Konrad Piascik, Ricardo Cabello, David Kim, Ruofei Du}
  
  \authornote{Both authors contribute equally. $^\dagger$ equal contributions, sorted alphabetically. \\$^\ddagger$Corresponding author and directional lead: Ruofei Du, me [at] duruofei [dot] com.}
  \authornotemark[3]
\affiliation{
    \country{\href{https://github.com/google/xrblocks}{\faGitSquare}  \url{https://github.com/google/xrblocks} \\ \href{https://xrblocks.github.io}{\faGithubSquare} \url{https://xrblocks.github.io} \\}
    \institution{\textbf{Google XR Labs}}
}
\renewcommand{\shortauthors}{Li et al.}

%% file: 0_abstract.tex
\begin{abstract}
We are on the cusp where Artificial Intelligence (AI) and Extended Reality (XR) are converging to unlock new paradigms of interactive computing. However, a significant gap exists between the ecosystems of these two fields: while AI research and development is accelerated by mature frameworks like JAX and benchmarks like LMArena, prototyping novel AI-driven XR interactions remains a high-friction process, often requiring practitioners to manually integrate disparate, low-level systems for perception, rendering, and interaction. To bridge this gap, we present \textsc{XR Blocks}, a cross-platform framework designed to accelerate human-centered AI + XR innovation. \textsc{XR Blocks} strives to provide a modular architecture with plug-and-play components for core abstraction in AI + XR: user, world, peers; interface, context, and agents. Crucially, it is designed with the mission of ``\textit{reducing frictions from idea to reality}'', thus accelerating rapid prototyping of AI + XR apps. Built upon accessible technologies ({WebXR}, {three.js}, {TensorFlow}, {Gemini}), our toolkit lowers the barrier to entry for XR creators. We demonstrate its utility through a set of open-source templates, samples, and advanced demos, empowering the community to quickly move from concept to interactive XR prototype.
\end{abstract}

%% file: _css.tex


%% file: _keywords.tex
\keywords{Extended Reality, Software Development Kit, WebXR, WebGL, Programming Language, Depth-based Interaction, Mixed Reality, Augmented Reality, Virtual Reality, Toolkit, AI, Gemini, Android XR, TensorFlow Lite, LiteRT}

%% file: 0_teaser.tex
\begin{teaserfigure}
  \centering
  \includegraphics[width=\textwidth]{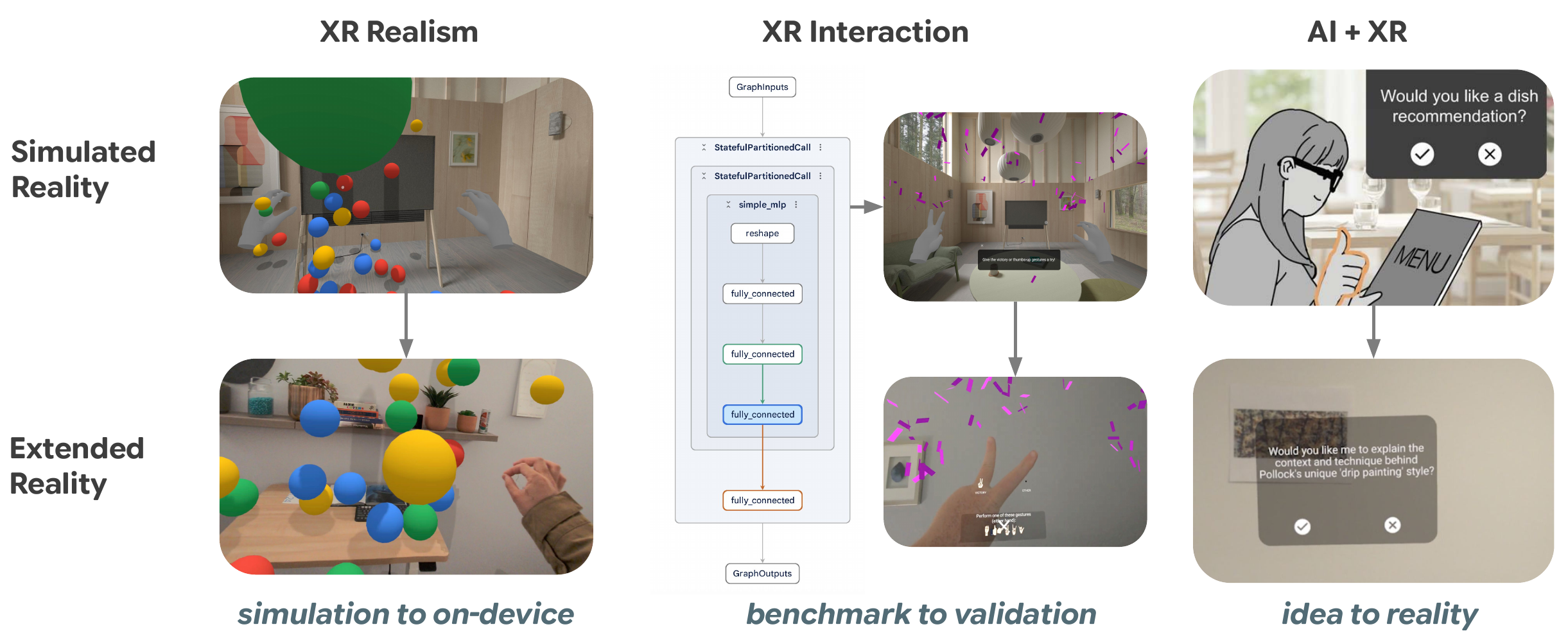}
  \caption{\systemname accelerates the prototyping of real-time AI + XR applications across desktop simulators and XR devices. Examples: (a) \textbf{XR Realism}: Prototype depth-aware, physics-based interactions \cite{Du2020DepthLab} in simulation and deploy the same code to real-world XR devices. (b) \textbf{XR Interactions}: Seamlessly integrate custom gesture models to desktop simulator and on-device XR deployment. (c) \textbf{AI + XR Integration}: Build intelligent, context-aware assistants, like the Sensible Agent \cite{Lee2025Sensible}. }
  \Description{}
  \label{fig:teaser}
\end{teaserfigure}

%% file: 1_introduction.tex
\section{Introduction}

The convergence of Artificial Intelligence (AI) and Extended Reality (XR) is reshaping human-computer interaction, unlocking innovation to empower human with augmented perception, augmented reasoning, and augmented interaction with the physical world, other people, devices, and AI agents.

The AI community has demonstrated the power of a robust ecosystem: frameworks like {PyTorch} \cite{Paszke2019PyTorch} and {JAX} \cite{Jax2018Github}, coupled with open benchmarks like ImageNet \cite{Deng2009ImageNet} and LMArena \cite{Chiang2024Chatbot}, have created a virtuous cycle of rapid, community-driven innovation. XR, however, has yet to tap into this flywheel effect. Its research and development landscape remains fragmented and fraught with friction, hindering the translation of innovative ideas into robust, interactive, on-device experiences. For example, algorithms prototyped in a specific game engine (Unity) six years ago in \textsc{DepthLab} \cite{Du2020DepthLab} are non-trivial to deploy on modern XR hardware. Oftentimes, creators are left to contend with the immense overhead of low-level systems programming, on-device testing, and endless one-off user studies just to validate a single interaction or real-time algorithm. While \textsc{MediaPipe} \cite{Lugaresi2019MediaPipe,Zhang2020Mediapipe} and \textsc{Visual Blocks} \cite{Du2023Rapsai,Zhou2023InstructPipe} offer novel solutions with coding and no-code frameworks to streamline AI pipeline development, their applications in XR are limited due to the missing access to sensors and natural user input.

This challenge is becoming more significant with the rise of a new paradigm--vibe coding \cite{Edwards2025Will}--an intent-driven creation. Platforms such as {Gemini Canvas} \cite{Google2025GeminiCanvas} and {Cursor} \cite{Cursor2025Cursor} provide a no-coding environment for anyone to author end-to-end mobile and web applications. While their 3D generation process are still rapidly advancing, we believe there will still be a critical gap preventing this magic from extending seamlessly into the XR domain. It is one thing to \textit{generate a 3D model of a frog-like hat} with core technologies like DreamFusion \cite{poole2022dreamfusion} and Magic3D \cite{lin2023magic3d}; it is another entirely to empower a user to simply \textit{pinch and summon a frog-like hat on the real-world desk, then hand over to remote participant}, as demonstrated in \texttt{Thing2Reality} \cite{Hu2025Thing2Reality}. This seemingly simple instruction requires a complex, implicit understanding of environmental perception, user intent, and atomic hand interactions--a specification that is currently missing. We envision a future where developers and designers specify novel interactions \textit{once}, using a high-level, human-centered language, and see them deployed \textit{everywhere}, including desktops, mobile, and XR headsets. They should be free to focus on inventing new user experiences, not re-implementing the fundamental mechanics of perception and interaction, where the individual perception modules are easily interchangeable.

To invent this future, we introduce \systemname, a framework engineered to accelerate human-centered AI + XR innovation. As presented in \autoref{fig:teaser}, \systemname offers an SDK to allow creators to simulate XR interaction directly on desktop and test them on device, benchmark a custom gesture model with virtual hands and then test it with real hand, and augment human abilities by fusing AI into XR. While \systemname is not the silver bullet for this grand challenge, but rather we create a proof-of-concept framework by distributed and part-time team members to provide the AI + XR community with the critical missing element: \textbf{a set of open source abstraction} for the core components of intelligent immersive systems, including user representation, environmental perception, and spatial interaction. It is architected to streamline the complex interplay between these elements, drastically reducing the code and expertise required to bridge the gap between concept to interactive prototype. Its initial version is built as \texttt{xrblocks.js}, a lightweight, cross-platform library on accessible web technologies (\texttt{WebXR} \cite{webxr}, \texttt{three.js} \cite{threejs}, \texttt{TensorFlow} \cite{Smilkov2019TensorFlow}, and \texttt{Gemini} \cite{team2023gemini}). We envision a future extension of \systemname will extend to native platform with LLM-powered compilers.

\begin{figure}[ht]
  \centering
  \includegraphics[width=\linewidth]{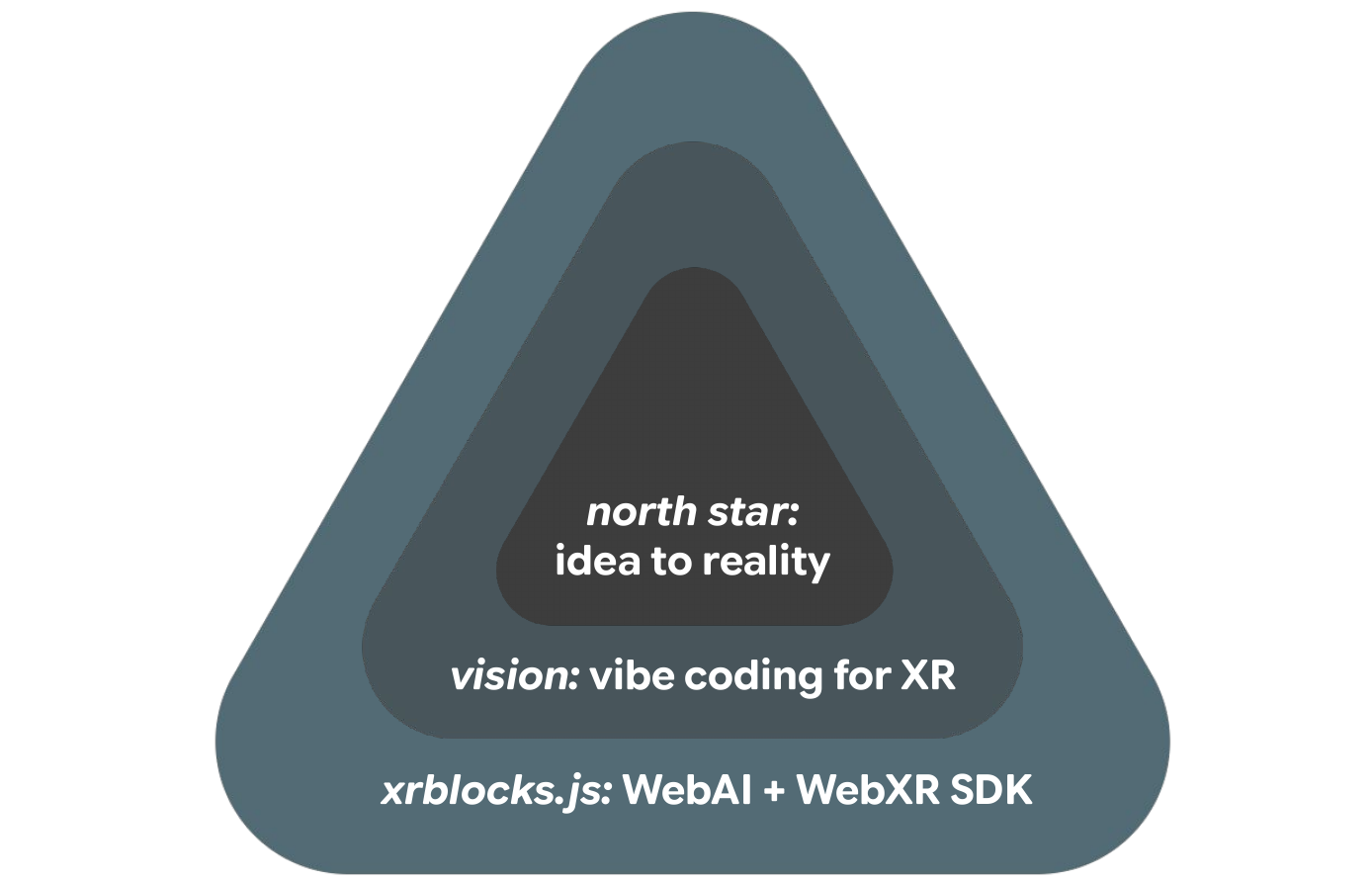}
  \caption{\systemname roadmap. This initial version of \texttt{xrblocks.js} serves as an opensource framework to accelerate prototyping with WebAI and WebXR. It should be iterated to empower vibe coding for XR. We envision this framework to achieve ``idea to reality'' in XR that follows human-centered objectives~\cite{Chen2023Next}: (i) aligning with human values in XR; (ii) assimilating human intents in XR; and  (iii) augmenting human abilities in XR.}
  \label{fig:vision}
\end{figure}
We demonstrate \systemname's power and flexibility through a suite of open-source toolkit at \url{https://github.com/google/xrblocks}, sample applications, and advanced demos that showcase tangible pathways for rapid prototyping, from depth-aware interaction, gestural controls, to dynamic generative human-AI interactions with a user's physical space. Not that this is a directional white paper in which some concepts marked in $^*$ have not yet landed in the SDK.






%% file: 2_related_work.tex
\section{Related Work}
\oursystem strives to offer a high-level abstraction and a set of tools for creators to author AI+XR experiences. Our work is built upon \texttt{three.js}, a mature WebGL rendering library, and is informed by a rich history of prior art in systems, toolkits, and programming frameworks for both AI and XR. We position \systemname in the context of three main areas: frameworks for 3D and XR interaction, the ``ecosystem effect'' that accelerates AI innovation, and prior efforts to bridge AI with interactive 3D worlds.

\subsection{Frameworks for 3D Interaction and Extended Reality}

The challenge of simplifying 3D and XR application development is not new. Seminal research systems in the 1990s laid the foundational groundwork for device abstraction and interaction management. VR Juggler \cite{Bierbaum2001VR} provided a powerful, reconfigurable data-flow architecture for CAVE-like \cite{cruz1992cave} virtual environments, while the MR Toolkit \cite{Shaw1993Decoupled} was among the first to formalize device-independent interaction techniques. The widespread adoption of ARToolKit \cite{Kato1999Marker} democratized augmented reality research by providing a robust, open-source solution for fiducial marker tracking, enabling a generation of researchers to experiment with AR interfaces.

In the modern XR era, this pursuit has continued, largely bifurcated between native mobile SDKs and comprehensive game engines. Mobile-specific SDKs like Apple's \texttt{ARKit} \cite{arkit} and Google's \texttt{ARCore} \cite{arcore} provide powerful, low-level access to core tracking technologies like SLAM, plane detection, and image tracking. While essential, these SDKs primarily offer a foundation for perception, leaving the implementation of high-level interaction patterns and application logic to the developer. On the other end of the spectrum, engines like Unity \cite{unity}, Unreal \cite{unreal}, and Godot \cite{godot}, have become the de-facto standard for commercial XR development, offering a high-ceiling environment with extensive tooling. However, their complexity presents a high floor for rapid prototyping, and integrating novel, external AI models can introduce significant friction. Worsestill, newer game engines may deprecate or remove older APIs, requiring \textbf{extensive code modifications} to use the updated equivalents.

To address the need for XR interactivity, several higher-level frameworks have been built upon Unity. In 2016, the Mixed Reality Toolkit (MRTK) \cite{mrtk}, designed for Microsoft HoloLens \cite{hololens}, provides creators with a vast library of production-grade UI controls, advanced interaction models (such as gaze-pinch indirect manipulation), and data binding systems for dynamic UI. Later in 2018, Unity's XR Interaction Toolkit (XRI) \cite{xri} offers a integral high-level, component-based system for creating VR and AR experiences. Further, community-driven framekworks such as MKRT3 \cite{mrtk3} and VRTK \cite{vrtk}, embraces XRI and the OpenXR standard, embodying a modular and performance-oriented design philosophy.

In constrast to the native ecosystem, \texttt{WebXR}, as a W3C standard, offers unparalleled accessibility and cross-platform compatibility for XR development on the open web. Libraries like \texttt{three.js} \cite{threejs} and \texttt{Babylon.js} \cite{babylonjs} offer developers  direct, programmatic control over the rendering pipeline, scene graph, and interactions. Frameworks like \texttt{A-Frame} \cite{AFrame} have successfully lowered the barrier to entry for \texttt{WebXR} by providing a declarative, entity-component-system (ECS) architecture based on HTML. \systemname shares this goal of accessibility across platforms, but differs in focus, prioritizing a high-level, user-centric and AI-driven interaction model in XR over a purely declarative scene graph. Our work builds directly on the rendering capabilities of \texttt{three.js}, but provides a distinct contribution in the \textit{abstraction layer} for orchestrating perception, interaction, and AI-driven behavior.

\subsection{The Ecosystem Effect in Artificial Intelligence}

A key motivation for our work is the ``flywheel effect'' observed in the AI research community. The unprecedented pace of innovation in AI can be attributed to a symbiotic ecosystem composed of three pillars. First, standardized and extensible frameworks like \texttt{TensorFlow} \cite{Smilkov2019TensorFlow} and \texttt{PyTorch} \cite{Paszke2019PyTorch} provided a common language and powerful abstractions like automatic differentiation, freeing researchers to focus on model architecture. Second, canonical datasets and benchmarks like \texttt{ImageNet} \cite{Deng2009ImageNet} and LMArena \cite{Chiang2024Chatbot} created shared goals and objective measures of progress, fostering healthy competition and collaboration. Finally, open model hubs like \texttt{Hugging Face} \cite{huggingface, shen2024hugginggpt} and \texttt{TensorFlow Hub} \cite{tensorflowhub} democratized access to state-of-the-art pretrained models, creating a culture of sharing and composition.

This virtuous cycle does not yet exist for interactive XR research. Even reproducing a piece of XR research on an updated game engine or new hardware requires lots of work. \systemname is designed as a step toward fostering such an ecosystem: all demos made with \systemname can be reproducible by hosting independently on the web. By providing a common, high-level framework and open-sourcing our examples, we aim to create a shared substrate upon which the community can build, compare, and distribute novel human-centered AI+XR interactions.

\subsection{Bridging AI and Interactive 3D Worlds}

The goal of integrating AI with interactive 3D environments has a long history, particularly in the context of training embodied agents. Simulation platforms such as AI2-THOR \cite{Kolve2017AI2} and Habitat \cite{Savva2019Habitat,Szot2021Habitat,Puig2023Habitat3} provide photorealistic and physically accurate environments for training agents in tasks like navigation and object manipulation. These systems, while powerful, are designed for the offline or non-real-time training of autonomous agents. Our work differs in its focus: \systemname is designed for creating real-time, human-in-the-loop interactive experiences for end-users, where the goal is often human-AI collaboration rather than pure agent autonomy.

Other research has explored using natural language to manipulate 3D worlds like LLMR \cite{DeLaTorre2024LLMR} and Thing2Reality \cite{Hu2025Thing2Reality}, allowing users to construct or edit scenes declaratively. \systemname builds directly on this legacy, but integrates modern LLMs as a central mechanism for interpreting user intent from a richer, multi-modal context that includes not just language, but also gesture and environmental sensing. In doing so, \systemname synthesizes these threads, taking inspiration from the interactivity of HCI frameworks and the goals of embodied AI to provide a unique tool focused on the rapid prototyping of human-centered AI + XR systems. 

%% file: 3_design_space.tex
\section{Design Principles}

The \systemname team strives to foster an ecosystem of Human-AI creation for perceptive XR experiences. Our architectural and API design choices are guided by three core principles:

\begin{enumerate}
    \item \textbf{Embrace Simplicity and Readability}: Inspired by the philosophy of Python's Zen \cite{peters2004zen}, we prioritize clean, human-readable abstractions. Our goal is that a developer's Script should read like a high-level description of the desired experience. Simple tasks should be simple to implement, and complex logic should remain explicit and understandable. This is critical for a framework intended for rapid prototyping by a diverse community of creators.
    \item \textbf{Prioritize the Creator Experience}: Our primary goal is to make authoring intelligent and perceptive XR applications as seamless as possible. We believe that creators should focus on the user experience, not on the low-level ``plumbing'' of sensor fusion, AI model integration, or cross-platform interaction logic. \systemname is designed to absorb this incidental complexity, providing powerful, ready-to-use primitives.
    \item \textbf{Pragmatism over Completeness}: We follow a design philosophy of pragmatism, akin to ``worse is better'' \cite{gabriel1991rise}. The fields of AI and XR are evolving at an immense speed. A comprehensive, complex framework that attempts to be perfect will be obsolete upon release. We favor a simple, modular, and adaptable architecture that is ``good enough'' for a wide range of applications. 
\end{enumerate}

\section{An Architectural Design for Generative Reality}

Designing the next generation of XR experiences --- interactive worlds that are context-aware, augmented, and deeply personal --- requires a paradigm shift in our authoring tools. We must move beyond managing low-level rendering and perception pipelines and toward sculpting interactions and intent. Drawing inspiration from \textsc{Visual Blocks for ML} \cite{Du2023Rapsai} and \textsc{InstructPipe} \cite{Zhou2023InstructPipe}, we designed \systemname around a core principle: \textbf{to provide a high-level, human-centered abstraction layer that separates the \textit{what} of an interaction} (the \texttt{Script}) \textbf{from the \textit{how} of its low-level implementation}. This architecture is our first step toward a future of \texttt{Generative Reality}.

\subsection{The \texttt{Reality Model}: A High-Level Abstraction of Reality}

\begin{figure}[ht]
  \centering
  \includegraphics[width=\linewidth]{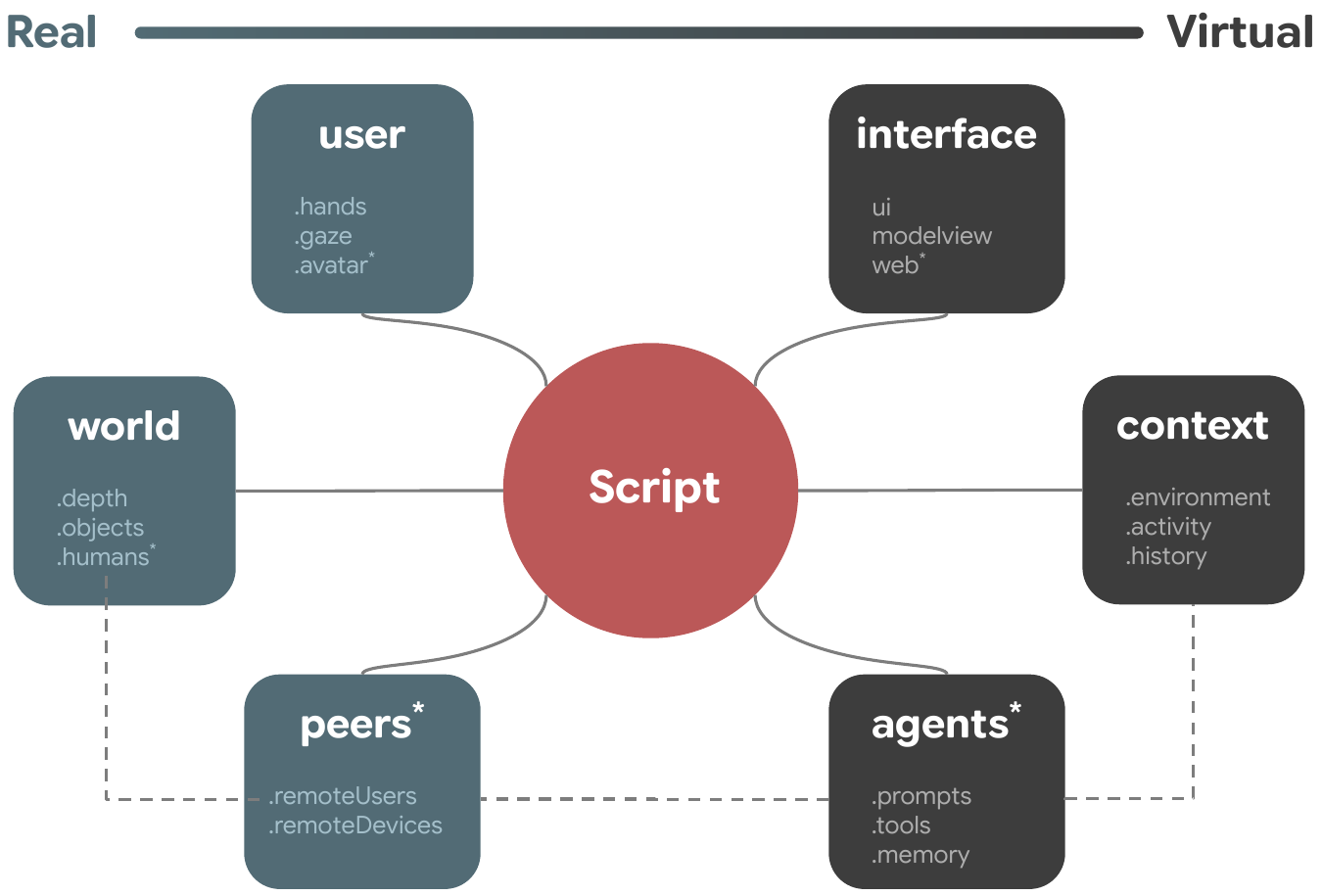}
  \caption{The conceptual \texttt{Reality Model} of the \systemname framework. At the center, the \texttt{Script} contains the application's logic and operates on a unified model of first-class primitives including the user, the physical world, AI agents, and the application context. Entities with $^*$ have not yet been fully implemented in the GitHub.}
  \label{fig:overview}
\end{figure}

At the heart of our design is \texttt{Script}, the narrative and logical center of an application. As shown in \autoref{fig:overview}, the \texttt{Script} does not operate on disconnected data streams, but on a unified \texttt{Reality Model}, a coherent, interactive representation of XR that elegantly fuses the physical and digital bits.

This model is composed of several first-class primitives:

\begin{itemize}
  \item \textbf{User \& the Physical World}: Our model is centered around the \texttt{User}, which is represented by their hands, gaze, and avatar. The physical \texttt{World} allows the \texttt{Script} to query the perceived reality such as depth, estimated lighting condition, objects, and tracked humans.
  
  \item \textbf{Virtual Interfaces \& Context}: The model augments the blended reality with virtual UI elements, from 2D panels to fully 3D assets. Perception pipeline analyzes the context of environment, activities, and histories of interaction.
  
  \item \textbf{Intelligent \& Social Entities}: AI-driven agents and remote human peers are not external components but are treated as primary entities within the model. This allows for rich, contextual interactions between all participants, real and artificial.
\end{itemize}

\begin{figure}[t]
  \centering
  \includegraphics[width=\linewidth]{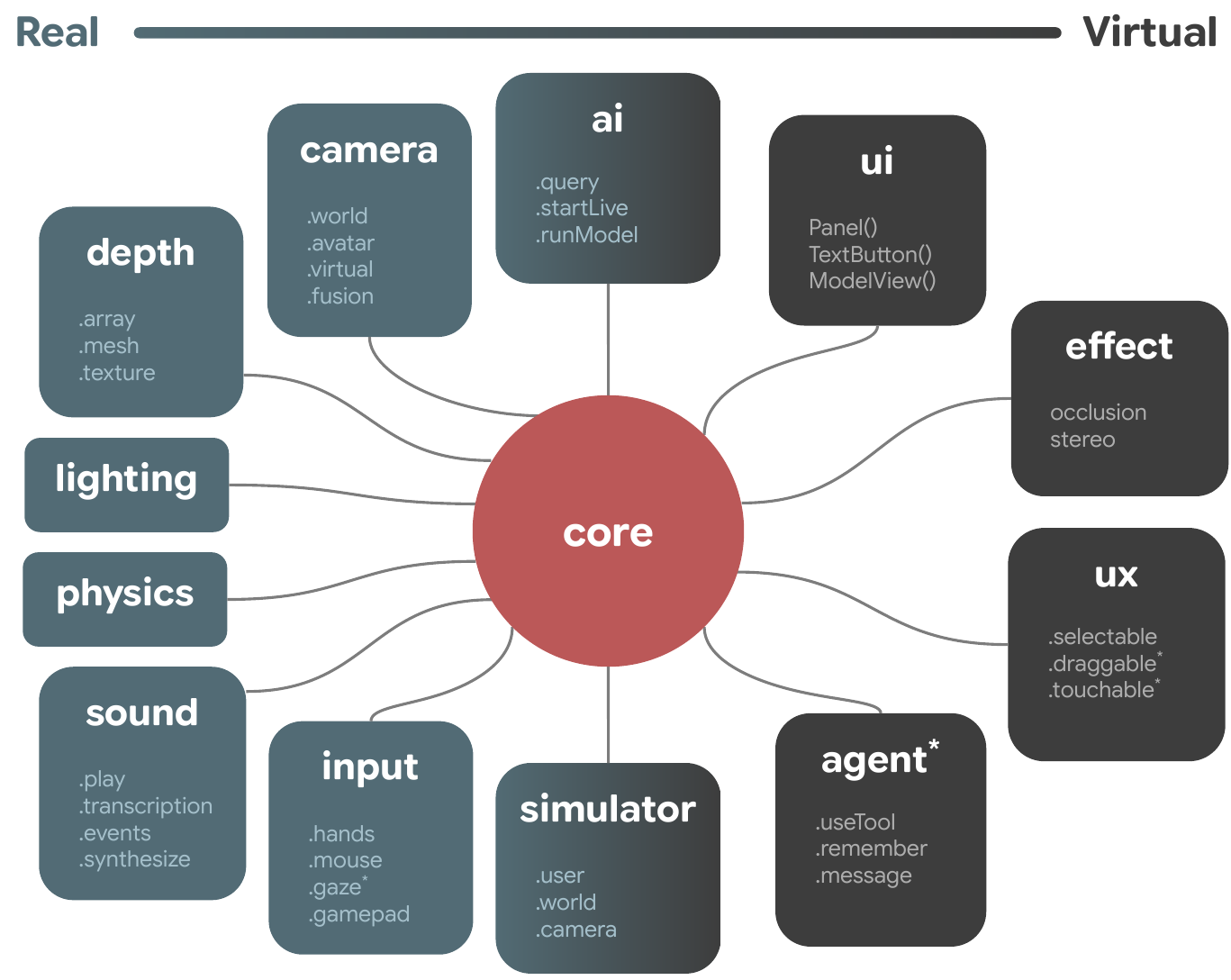}
  \caption{The modular architecture of the \systemname's \texttt{Script} engine. The \texttt{core} engine consists of essential subsystems to realize the framework's high-level abstractions, spanning perception (\texttt{depth}, \texttt{input}), AI integration (\texttt{ai}, \texttt{agent}), and user experience (\texttt{ui}, \texttt{ux}). Subsystems with $^*$ have not yet been fully implemented in the GitHub. While we show a few examples in XR Blocks, many implementations are far from perfection.}
  \label{fig:subsystems}
\end{figure}

\subsection{The \texttt{Core} Engine: An Extensible Toolkit for Authoring AI + XR Experiences}

This high-level \texttt{Reality Model} is realized by \systemname's modular Core engine (\autoref{fig:subsystems}), an extensible toolkit that orchestrates the complex subsystems required for spatial computing. The Core provides a high-level API that enables developers to harness these subsystems without needing to master the implementation.

\begin{figure}[h]
  \centering
  \includegraphics[width=\linewidth]{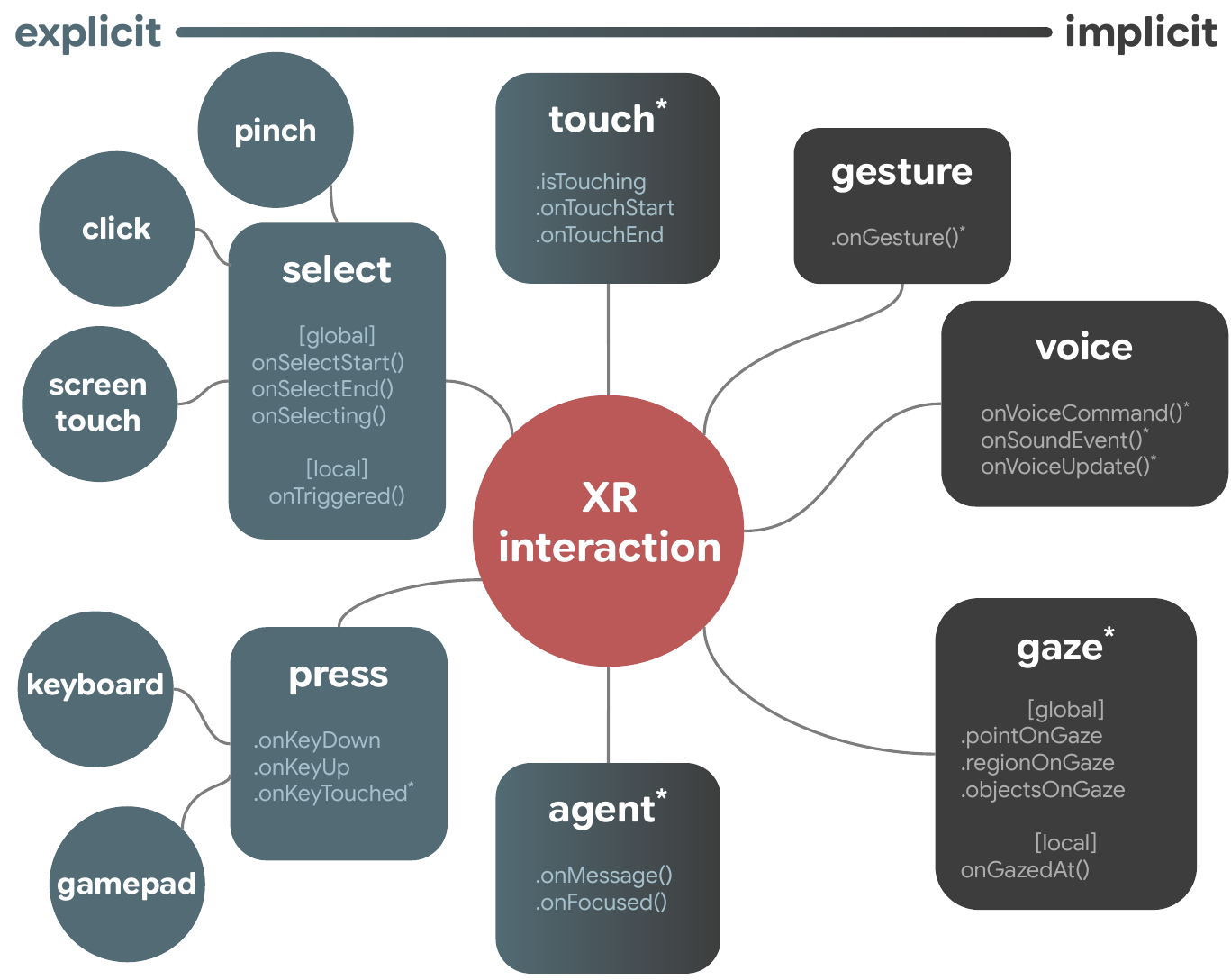}
  \caption{The Interaction Grammar of \systemname, which abstracts user input by distinguishing between two types of interaction. Explicit events are direct, low-level inputs (\textit{e.g.}, a touch or click), while implicit intents are higher-level interpretations (\textit{e.g.}, a gesture or voice command), allowing creators to build interaction against user intent.}
  \label{fig:interaction}
\end{figure}

\begin{itemize}
  \item \textbf{Perception \& Input Pipeline} The \texttt{camera}, \texttt{depth}, and \texttt{sound} modules continuously feed and update the \texttt{Reality Model}'s representation of physical reality. The \texttt{input} module normalizes user actions from various devices, providing the raw data for the Interaction Grammar to interpret.
  
  \item \textbf{AI as a Core Utility} The \texttt{ai} module acts as a central nervous system, providing simple yet powerful functions (\texttt{.query}, \texttt{.runModel}) that make large models an accessible utility. This is complemented by the \texttt{agent} module, which provides AI entities with concrete capabilities to act and remember (\texttt{.useTool}, \texttt{.remember}).

  \item \textbf{Experience \& Visualization Toolkit} To enable rapid creation, the toolkit provides a library of common affordances. The \texttt{ux} module offers reusable interaction behaviors like \texttt{.selectable} and \texttt{.draggable}, while the \texttt{ui} and \texttt{effect} modules handle the rendering of interfaces and complex visual effects like occlusion.

\end{itemize}


  
  
  
  
  

\subsection{A Vision for AI + XR Creative Prototyping}

By separating the abstract Reality Model from the concrete Core engine, \systemname enables a powerful new creative workflow. It allows creators to move from high-level, human-centric ideas to interactive prototypes at unprecedented speeds. We envision a future, where a declarative prompt, \textit{``When the user pinches at an object, an agent should generate a poem of it.''}, could be directly translated to high-level instructions in \systemname:

\begin{lstlisting} [basicstyle=\footnotesize]
  for (const object of world.objects) {
    if (user.isSelectingAt(object)) {
      const prompt = 'Write a poem with ${object.name}.';
      const poem = agent.query(prompt);
      this.textView.setText(poem);
    }
  }
\end{lstlisting}

\begin{figure*}[t]
  \centering
  \includegraphics[width=0.9\linewidth]{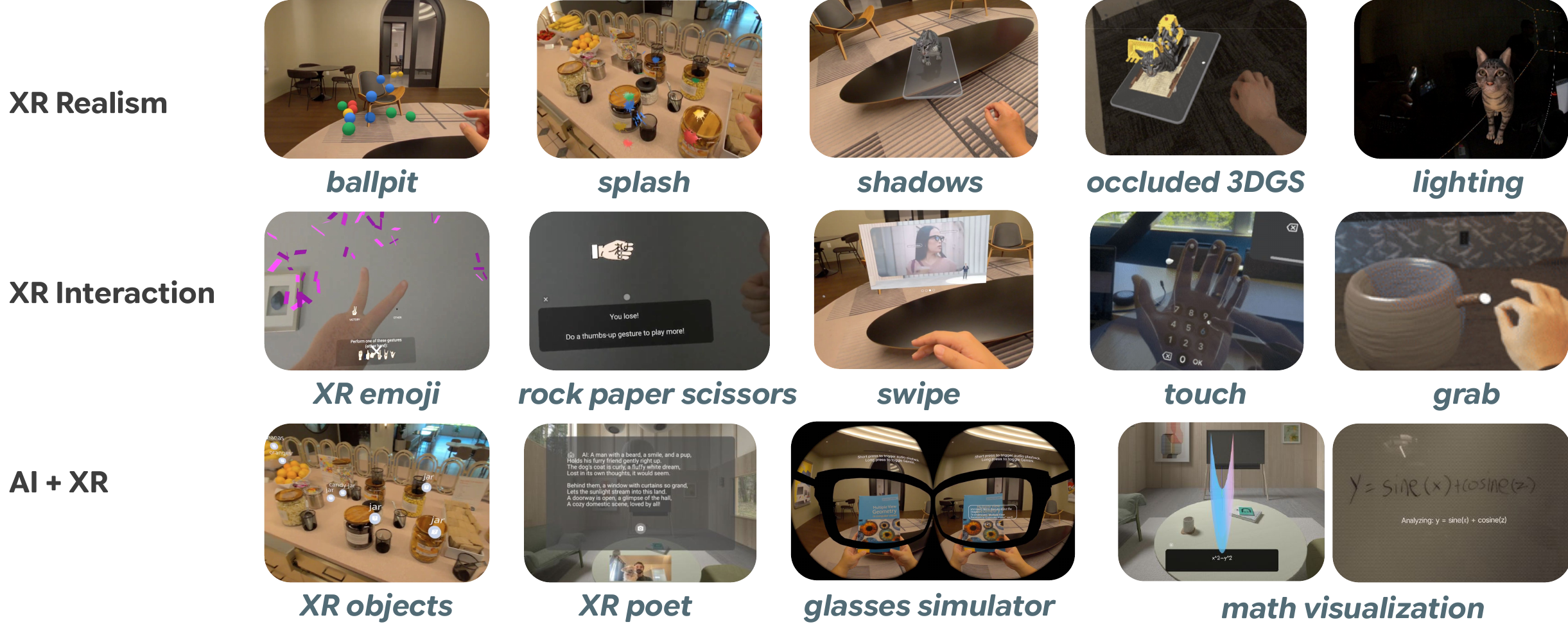}
  \caption{Applications of \systemname. (1) \textbf{XR Realism}: depth-aware and physics-based ballpit and splash games; geometry-aware shadows, 3D Gaussian splatting with occlusion, and lighting estimation. (2) \textbf{XR Interaction}: immersive emoji and rock paper scissors game empowered by custom ML models, dynamic swipe recognition, touch and grab with the physical world. (3) \textbf{AI + XR}: integration with Gemini Live, XR objects, glasses simulation in XR, and poem generation with real-world camera.}
  \label{fig:applications}
\end{figure*}

Hence, creator's prompt is no longer a pseudocode but a direct summary of the implementation logic. This envisioned framework seamlessly translates this intent into a system-level execution flow, composing capabilities from the \texttt{input}, \texttt{sound}, \texttt{ai}, \texttt{world}, \texttt{ui}, and \texttt{agent} modules to generate an emergent, intelligent behavior.

This is merely the first step. The architecture of \systemname points toward a future where generative AI is applied not just to conversational agents, but to the fabric of reality itself. Future work can explore scripting with even higher-level intents, such as, \textit{``Make this room feel like a cyberpunk cafe''} allowing the \texttt{depth}, \texttt{lighting}, \texttt{sound}, and \texttt{agent} modules to collaborate in synthesizing a complete, multi-sensory experience. This opens avenues for differentiable world-building, or world-building with LLM-driven ``Gradient'' inspired by TextGrad \cite{Yuksekgonul2024TextGrad} and ToolGrad \cite{Zhou2025ToolGrad}, where environmental aesthetics and even procedural geometry could be optimized by AI in response to real-world context or user emotion. Our framework aims to provide a substrate for exploring learnable interaction grammars, where systems could co-evolve personalized and more expressive communication modalities with their users over time.

Ultimately, \systemname is not just a toolkit but a hypothesis about the future of creative expression: a future where the boundary between programming, design, and conversation dissolves, enabling everyone to script realities as fluidly as we script stories.



%% file: 5_applications.tex
\section{\systemname Applications}

To demonstrate the expressive power and flexibility of the \systemname architecture, we developed a suite of demonstrative applications and interactive examples in \url{https://xrblocks.github.io/docs/samples/ModelViewer/}. These are not merely tech demos but tangible explorations into future interactive paradigms, showcasing how our framework's core principles enable the rapid prototyping of experiences that were previously complex and costly to build. The following examples highlight how \systemname facilitates the creation of realistic, interactive, and intelligent mixed-reality worlds.

\subsection{Enhancing XR Realism}

A fundamental requirement for immersive experiences is the seamless blending of real and virtual elements. \systemname's unified \texttt{Reality Model} allows creators to achieve this with minimal effort by composing high-level modules.

\textbf{Dynamic Occlusion and Relighting}: By leveraging the depth \cite{valentin2019depth} and lighting estimation \cite{legendre2019deeplight} modules, virtual objects are rendered with a deep awareness of the physical environment. A person can realistically walk behind a real-world sofa to observe a virtual cat, and a physical light source can cast light and shadows that interact correctly with the virtual cat and the room's actual geometry. This complex behavior is achieved with a few declarative lines, as \systemname handles the continuous fusion of real-world sensor data from WebXR API with the virtual scene's rendering pipeline.

\textbf{Physics-Based World Interaction}: We demonstrate how the physics module can be applied to virtual objects \href{https://xrblocks.github.io/docs/samples/Ballpit}{here}, allowing them to interact with the real world's geometry \cite{izadi2011kinectfusion,Du2020DepthLab}. For example, a creator can enable depth and physics in \systemname and instantiate a virtual ball in \texttt{onSelect()}. When a user shoots the ball, it realistically bounces off real-world geometries, with its trajectory and behavior governed by the live depth map of the room.



\subsection{Intelligent Environments with AI + XR Integration}

The true power of our framework is realized when the \texttt{Reality Model} is deeply integrated with generative AI. This allows for the creation of dynamic, personalized environments that respond intelligently to user intent.

\textbf{Augmented Object Intelligence} (\textsc{XR-Objects}): Building on our prior work \cite{Dogan2024Augmented}, we use \systemname to imbue everyday physical objects with interactive, digital affordances in \href{https://xrblocks.github.io/docs/samples/Gemini-XRObject}{this demo}. When a user performs a long pinch, the \texttt{ai} module identifies the objects in the physical world. The framework then dynamically attaches a set of virtual buttons and allows users to touch these buttons to ask questions. This envisions a future of turning passive physical objects into active, programmable interfaces.

\textbf{Proactive and Unobtrusive Assistance}: \systemname served as the foundation for \textsc{Sensible Agent} \cite{Lee2025Sensible}, a system for proactive and unobtrusive interaction with an AR assistant. \textsc{Sensible Agent} tackles the critical challenge that many AR agents are intrusive and distracting, proposing a framework where agents only intervene when it is appropriate and helpful. 
Developing this system was significantly accelerated by \systemname. The agent's ability to perceive the user's current activity and environment is directly supported by our unified \texttt{Reality Model}. Its core logic for deciding when to intervene was implemented using the \texttt{ai} module, while the design and testing of its subtle, peripheral notifications were rapidly prototyped with the \texttt{ui} modules. Ultimately, \textsc{Sensible Agent} is a powerful testament to our framework's primary goal: by providing robust, high-level tools for perception and interaction, \systemname empowers HCI researchers to focus on higher-order challenges, such as the social and cognitive principles of human-agent collaboration.


%% file: 7_future_work.tex
\section{Discussion}

By abstracting the immense complexity of underlying hardware and AI models, \systemname envisions a future where creators and AI can operate at the speed of thought. The core contribution is not merely a new library, but an argument for a fundamental shift in how we approach the authoring of interactive systems: a shift from low-level programming to high-level, intent-driven creation.

\subsection{From Language Models to AI+XR Models}

The rapid commoditization of large language models has unlocked unprecedented generative capabilities. However, a model that can describe a ``cyberpunk cafe'' in text with 3D diffusion models is fundamentally different from a system that can render one in a user's physical space, complete with appropriate lighting, sound, and interactive agents. The challenge lies in grounding generative AI in the rich, multi-modal context of a user's reality. \systemname is our first step toward bridging this gap. Our ``\texttt{Reality Model}'' provides the necessary structure to channel the raw potential of generative AI, transforming it from a disembodied ``chatbot'' into an embodied, context-aware collaborator capable of perceiving, understanding, and modifying a shared reality.

\subsection{Limitations and Future Work}

\systemname provides a robust architectural blueprint for a new era of AI + XR development. Our current implementation prioritizes the core abstractions, and its limitations highlight several exciting avenues for future research.

\textbf{System design $\textit{vs.}$ implementation}: 
Our framework's conceptual primitives—agents, peers, and context—are depicted with a starting point, yet not fully implemented. The agent primitive, for instance, could be extended to support richer personalities, persistent memory, and more sophisticated proactive behaviors. Similarly, the future peers model should lay the groundwork for complex cross-device interactions that go beyond simple data sharing \cite{Zhu2025Beyond}. A critical next step is to enhance the context and user models to perceive and adapt to Situationally Induced Impairments and Disabilities (SIIDs) \cite{Liu2024Human}, making AI-driven assistance truly accessible and equitable.

\textbf{Performance and Latency}: Our choice of web technologies prioritizes accessibility but brings known trade-offs. The framework can never match the rendering performance of native engines like Unity or Unreal, and reliance on cloud AI models introduces network latency. Near-term future work will explore hybrid architectures that combine our framework's flexibility with native performance, alongside on-device model distillation for real-time AI. A more visionary goal is to develop \textit{an LLM-driven cross-compiler} capable of translating a high-level \systemname script into optimized, native code for multiple target engines, effectively solving the trade-off between ease-of-use and raw performance.

\textbf{The Abstraction Ceiling}: 
Our high-level abstractions, by design, prioritize rapid prototyping over ultimate expressivity. This ``abstraction ceiling'' means there will always be novel or performance-critical interactions that require lower-level control. A key future goal is to design an elegant ``escape hatch'' mechanism. This would allow expert developers to seamlessly write custom shaders, interface directly with device sensors, or implement bespoke interaction logic, all without breaking the integrity of the high-level \systemname framework.

\textbf{WebXR and Privacy}:
We chose WebXR to maximize accessibility and efficiency, but this comes with a trade-off: to protect user privacy, web standards deliberately restrict access to sensitive sensor data like raw eye-tracking or face-tracking feeds, which are available in native OpenXR environments. We envision future browser-level APIs that allow for privacy-preserving on-device processing. In this model, raw sensor data would never leave the user's device. Instead, the browser may expose a secure, sandboxed API providing high-level, privacy-safe signals. This requires more resources and efforts from both industry and community.

\subsection{The Future of XR Interaction and Creation}

Our architecture opens the door to several exciting, long-term research directions.

\textbf{Learnable Interaction Grammars}: Currently, our mapping from implicit intents  to explicit events is hand-designed. The \systemname framework provides the ideal substrate to explore systems that could learn a user's unique interaction preferences and gestural vocabulary over time, leading to truly personalized and co-adaptive interfaces. Future endeavors shall graudually extend seminal works in gestures \cite{Wang2021GesturAR,Pei2022Hand}, locomotion methods \cite{Di2021Locomotion}, and cross-device interaction \cite{Brudy2019Cross,Zhu2025Beyond} as primitives into the SDK to allow for a richer library towards learnable interaction grammar.

\textbf{Differentiable and Co-Adaptive Realities}: For the graphics and AI communities, our framework suggests a future where entire scenes and agent behaviors are procedurally optimized. If a user states, ``make this room feel more relaxing'', a differentiable rendering pipeline could be guided by an AI model to tune lighting, color, and even geometry to achieve the desired affective goal. This same principle extends to the agents within a mirrored world \cite{Du2016Social,Du2019Geollery}. An agent's behaviors, conversational style, and animations could be co-optimized within this loop to better suit the user's personality and context. Realizing this vision will require deep integration between interactive platforms like \systemname and modern ML frameworks, a challenge our modular architecture is designed to facilitate.

\textbf{Multi-Sensory Synthesis}:
We have focused on audio-visual experiences, but the \systemname model is designed to be extensible. Integrating modules for haptics, EEG, and other sensory modalities could allow creators to compose truly immersive, multi-sensory narratives. E.g., creators today can extend \systemname with Arduino through WebUSB protocol.


%% file: 8_conclusion.tex
\section{Conclusion}

The process of creating intelligent, interactive XR experiences is currently too fragmented and complex, placing a significant barrier between a creator's vision and its realization. In this paper, we presented \systemname, an architecture and toolkit designed to dissolve this complexity. By providing a high-level, human-centered abstraction layer that separates the \textit{what} of an interaction from the \textit{how} of its low-level implementation, \systemname dramatically accelerates the prototyping process. We have demonstrated how this approach enables the rapid development of sophisticated applications that seamlessly blend the real and virtual, and are imbued with contextual intelligence. \systemname is more than a toolkit; it is a foundational step toward a future where the boundary between programming, design, and conversation disappears, enabling us to script realities as fluidly as we script stories. However, \systemname is far from completion and this white paper serves as an initial visionary documentation to attract more creators to join our journey. We believe: with the right set of tools, everyone can unleash their inner creativity with AI.


\begin{acks}
We would like to deeply thank Xiuxiu Yuan, Oren Haskins, Brian Collins, Barak Moshe, Seeyam Qiu, Steve Toh for the UX Design Support; Yinghua Yang and Brenton Simpson for the cross-team collaboration; Tim Bettridge, Kevin Lam for significant contributions to XR Blocks x Gemini Canvas; Furthermore, we would like to extend thanks to the seminal work in Android XR Unity Samples with Mahdi Tayarani, Max Dzitsiuk, Patrick Hackett, all of DepthLab~\cite{Du2020DepthLab}, Ad hoc UI~\cite{Du2022Opportunistic}, Rapsai (Visual Blocks)~\cite{Du2023Rapsai}, InstructPipe~\cite{Zhou2023InstructPipe}, DialogueLab~\cite{Hu2025DialogLab}, and Sensible Agent~\cite{Lee2025Sensible} co-authors, which greatly inspired this project along the way; we further extend thanks to Eric Gonzalez, Nicolás Peña Moreno, Max Spear, Yi-Fei Li, Ziyi Liu, Jing Jin, Adarsh Kowdle, and Guru Somadder for their insightful feedback and support in XR Blocks.
\end{acks}